\documentclass[preprint2]{aastex} 

\shortauthors{Fassnacht \& Lubin}
\shorttitle{A Group Coincident with CLASS B0712+472}

\begin{document}

\title{The Gravitational Lens -- Galaxy Group Connection:
I. Discovery of a Group Coincident with CLASS B0712+472}

\author{C. D. Fassnacht}
\affil{Space Telescope Science Institute, \\
3700 San Martin Drive, Baltimore, MD 21218}
\email{cdf@stsci.edu}

\author{L. M. Lubin\altaffilmark{1}} 
\affil{Department of Physics and Astronomy, \\
Johns Hopkins University, Baltimore, MD 21218}
\email{lml@stsci.edu}

\altaffiltext{1}{Current address: Space Telescope Science Institute,
3700 San Martin Drive, Baltimore, MD 21218}

\begin{abstract}

Previous observations of the environments of the lensing galaxies in
gravitational lens systems suggest that many of the lensing galaxies
are associated with small groups of galaxies. As a result, we have
begun a coordinated program to study the local environments of all
known gravitational lens systems.  In this paper, we present results
on the gravitational lens system CLASS B0712+472, which has previously
measured source and lens redshifts of $(z_{\ell},z_s) =
(0.4060,1.339)$.  Although we have not found a galaxy group associated
with the primary lensing galaxy, we have found a foreground group
which is spatially coincident with the lens system. Based on
multi-object spectroscopy taken at the Keck 10-m telescope, we have
confirmed ten group members with a mean redshift of $\bar{z} =
0.2909$. The resulting velocity dispersion and estimated virial mass
are $306^{+110}_{-58}$ km s$^{-1}$ and $3.0^{+2.2}_{-1.2} \times
10^{13}~h^{-1} M_{\odot}$, respectively, for
$(\Omega_m,\Omega_\Lambda) = (0.2,0.0)$. The dynamical properties of
this moderate-redshift group are completely consistent with the range
of values found in nearby groups of galaxies.  Five of the group
members are red, elliptical-like galaxies, while the remaining five
are active, star-forming galaxies.  Based on the spectroscopic results
and the publically-available {\it Hubble Space Telescope} imaging of
nine group members, we find that the early-type fraction is 40\%. We
estimate that the effect of this foreground group on the gravitational
lensing potential of B0712+472 is small, producing an external shear
which is only a few percent, although the shear could be larger if the
group centroid is significantly closer to the lens system than it
appears to be.

\end{abstract} 


\section{Introduction}

Studies of the large scale structure in the local universe indicate
that a significant fraction of galaxies reside in groups.  Redshift
surveys out to $\sim 15,000\ {\rm km~s^{-1}}$ find that 50 -- 60\% of
galaxies are members of groups, with the most common systems being
small, loose groups with 4 to 5 galaxies
\citep{tg76,geller83,tully87,ramella89}.  As such, small groups of
galaxies make a substantial contribution to the total mass of the
universe and constitute the most common galaxy environment.  Groups of
galaxies have been well-studied at low redshifts
\citep[e.g.,][]{tully87,dellantonio94,mulchaey96,mahdavi97,zm98}.
However, little work has been done on more distant groups of galaxies
because the modest optical overdensities of small groups make them
difficult to detect at intermediate and high redshift.

Recent data suggest that gravitational lens systems may provide a
method for finding moderate and high-redshift groups of galaxies.
Specifically, observations of four gravitational lens systems
(MG~0751+2716, PG~1115+080, MG~1131+0456, and JVAS~B1422+231) have
detected and spectroscopically confirmed groups of galaxies which are
associated with the primary lensing galaxy
\citep{kundic1115,kundic1422,tonry1422,tonry0751,tonry1131}.  In
addition, there are at least two lens systems (CLASS~B1359+154 and a
second group for MG~1131+0456) for which the high spatial density and
colors of the galaxies close to the lens system suggest the presence of a
group \citep{rusin1359,kochanek00}.  All of these groups are compact,
with at least three members within a projected radius of
$\sim$30\arcsec\ (within 1\farcs7 in the case of B1359+154).  We note
that several gravitational lenses are associated with more massive
systems, namely clusters of galaxies.  The most famous example is the
first gravitational lens to be discovered Q0957+561
\citep{0957discovery,0957cluster}.  However, the project described in
this paper is concerned only with small groups of galaxies.

Although these data are not conclusive, they strongly suggest that the
primary lensing galaxies in gravitational lens systems are associated
with groups of galaxies, implying that a larger association of
galaxies may contribute to the lensing mass distribution.  In
addition, \citet{bsk01} analyze the lensing rates in the Hubble Deep
Field North \citep[HDF-N;][]{hdfn} and find that the number of
multiple-imaged galaxies is an order of magnitude smaller than
predicted on the basis of radio lens surveys.  They argue
that this strong deficit of gravitational lenses in the HDF-N
\citep{zepf97,rdb98} is associated with the subsequent lack of
elliptical--group combinations at $z_{\ell} \sim 0.5$, the most likely
redshift for lensing. Hence, they postulate that most galaxy lenses
are located in compact groups.  Similarly, an analysis based on the
relationship between galaxy types and their local environment by
\citet{keetongrp} suggests that at least 25\% of gravitational lenses
should be found in groups or clusters.

Searches for gravitational lens are biased toward high mass systems
because these systems have a larger cross-section for lensing.
Consequently, theoretical analyses predict that most gravitational
lenses will be early-type (elliptical or S0) galaxies
\citep[e.g.,][]{tog,ft91,maozrix93,csk93} because these galaxies are
more massive than spirals.  Some spiral galaxy lenses have been
detected \citep[e.g.,][]{carilli0218,0712spiral,disklenses}.  However,
high angular-resolution images from the {\em Hubble Space Telescope}\
(HST) show that most lensing galaxies have properties consistent with
those of early-types \citep[e.g.,][]{keetongalevol,csk_fp}.  In the
nearby universe early-type galaxies are preferentially found in groups
or clusters \citep[e.g.,][]{dressler80,zm98}.  If this correlation
continues to higher redshift, a large fraction of gravitational lenses
may be found in groups of galaxies.  The mean redshift for
gravitational lenses is $\langle z_{\ell}\rangle \sim 0.6$.  Thus, if
the lens--group association is valid, gravitational lens systems can
be used to study groups of galaxies in a redshift range which has been
largely unexplored.  In addition, the major source of uncertainty in
$H_{0}$ measurements from gravitational lenses is the modeling of the
lensing potential \citep[e.g.,][]{1608mod1}.  If galaxy groups
contribute significantly to the lensing mass distribution or to an
external shear, the group mass needs to be measured.  A spectroscopic
survey and the resulting velocity dispersion measurements are the most
practical way to achieve this end.

In light of this, we have begun a coordinated effort to study the link
between groups of galaxies and gravitational lenses.  Specifically, we
are obtaining photometric and spectroscopic data on the fields
surrounding all lens systems where both the source and lens redshift
are currently known.  Our ground-based program is complemented by the
extensive set of gravitational lens images in the HST archive.  Many
of the archival images were obtained by CfA-Arizona Space Telescope
Lens Survey team \citep[CASTLES\footnote{See 
\url{http://cfa-www.harvard.edu/castles}};][]{castles1,castles2}, which is
conducting a program to obtain uniform multi-band ($V$, $I$, and $H$)
HST images of all known lens systems.

In this paper, we present the results from observations of the field
surrounding the gravitational lens system CLASS B0712+472.  Unless
otherwise noted, we use $h$ to express the value of the Hubble Constant,
where $H_{0} = 100~h~{\rm km~s^{-1}~Mpc^{-1}}$, and assume 
$\Omega_{m} = 0.2$, and $\Omega_{\Lambda} = 0.0$.

\section{CLASS B0712+472}

The B0712+472 gravitational lens system was discovered by
\citet{0712disc} as part of the Cosmic Lens All-Sky Survey
\citep[CLASS;][]{class_bos_1,class_bos_2}.  The maximum separation of
the four lensed images is 1\farcs{27}, and the flux density ratios at
15 GHz are $14.2:10.5:5.4:1$ for components ${\rm A:B:C:D}$,
respectively.  High-resolution radio maps from the Very Large Array
(VLA) and the Multiple-Element Radio-Linked Interferometer (MERLIN) do
not resolve the four images \citep{0712disc}.  Images taken with the
Wide Field Planetary Camera 2 (WPFC2) aboard HST show the three
brightest images of the background source; in addition, the lensing
galaxy is detected with optical magnitudes of $V \sim 22.2$ and $I
\sim 20.0$ in an elliptical aperture with major and minor axes of
2\arcsec\ and 1\arcsec, respectively \citep{0712disc}.  Images
obtained with the Near-Infrared Camera/Multi-Object Spectrograph
(NICMOS) aboard HST are similar to the WFPC2 images but show,
additionally, the fourth lensed image (image D) and a faint arc
produced by the host galaxy of the background quasar
\citep{0712nicmos}.

Preliminary spectroscopic observations of lens system were taken at
the William Herschel Telescope \citep{0712disc}; however, the definitive
measurement of the system redshifts were made by \citet{zclass}
using the Low Resolution Imaging Spectrograph \citep[LRIS;][]{lris}
on the Keck telescopes.  The resulting source and lens
redshifts are $(z_{\ell},z_s) = (0.4060, 1.339)$.  The background
source exhibits broad \ion{C}{3}] and \ion{Mg}{2} emission lines
typical of a quasar spectrum. The lensing galaxy has an early-type
spectrum which is characterized by a moderately strong 4000\AA\
break, small equivalent width Balmer absorption lines, and no
[\ion{O}{2}] emission.  The observed image splitting in B0712+472
implies that the mass and mass-to-light of the lensing galaxy within
the Einstein ring radius are $5.40 \pm 0.22 \times 10^{10}~h^{-1}~{\rm
M_{\odot}}$ and $8.6 \pm 0.9~h~{\rm (M/L)_{\odot}}$, respectively, for
$q_{0} = 0.5$ \citep{zclass}.

\section{Observations}

\subsection{Broad-Band Optical}

\subsubsection{Palomar 60--in}

Preliminary photometric observations were taken on January 3-4 and
April 1-2, 2000 with the CCD camera on the Palomar 60--in (1.5--m)
telescope.  The camera covers a field-of-view of $12\farcm9 \times
12\farcm9$.  The field containing the lens system was imaged in the
three filters, Gunn $g$, $r$, and $i$. The total exposure times were
$\{7200~3600~2400\}$ sec in $\{g~r~i\}$, respectively.  The average
seeing during these observations was 1\farcs{5}.  We calibrated the
Palomar images to the Gunn photometric system through exposures on
several Gunn standards \citep{gunnstd}. The typical variations about
the nightly photometric transformations are 0.08~mag or less.

Object detection, cataloging, and photometry were performed on the
final images using the software package SExtractor \citep{sextractor}.
We adopted a fixed detection threshold of 1.5-$\sigma_{\rm sky}$. This
threshold corresponds to a surface brightness of approximately
$\{\mu_{g}~\mu_{r}~\mu_{i}\} \approx \{26.1~25.8~25.1\}~{\rm
mag~arcsec^{-2}}$. A Gaussian filter with a FWHM of {1}\farcs{1} and a
minimum detection area of $1.4~{\rm arcsec^{2}}$ were used. The
objects detected in each image were visually inspected, so that
spurious detections (e.g., cosmic rays or diffraction spikes) could be
removed from the catalog. The number of spurious detections
corresponded to 10--15\% of all detections.  The resulting object
catalog contains 996 objects.  For each object, we have measured a
total magnitude, an aperture magnitude within a circular aperture of
radius 3\arcsec, and a star--galaxy classification
\citep[see][]{sextractor}.  We reach approximate limiting magnitudes
of $g \sim 23.0$, $r \sim 23.0$, $i \sim 22.0$ for a 5-$\sigma$
detection in our standard circular aperture.

\subsubsection{Keck 10--m}

Higher-quality photometric observations were obtained on January
17-19, 2001 with LRIS on the Keck I telescope. LRIS covers a
field-of-view of $6\arcmin \times 8\arcmin$. Data were taken in three
broad band filters, $BRI$, which match the Cousins system well.  The
total exposure times were $\{3000~1800~2400\}$ sec in $\{B~R~I\}$,
respectively. The Keck observations were calibrated to the standard
Cousins-Bessell-Landolt (Cape) system through exposures on a number of
Landolt standard star fields \citep{landoltstd}.  As in the processing
of the Palomar 60--in data, the SExtractor package was used to detect,
classify, and obtain aperture and total magnitudes for all objects in
the co-added $BRI$ images.  The detection thresholds in the three
bands were $\{\mu_{B}~\mu_{R}~\mu_{I}\} \approx
\{27.6~26.4~26.0\}~{\rm mag~arcsec^{-2}}$.

For the color analysis presented in this paper, we use aperture
magnitudes computed in a circular aperture with a radius of $3''$ from
these Keck observations.  This corresponds to a physical radius of
$10.5~h^{-1}~{\rm kpc}$ at $z = 0.4060$, the redshift of the lensing
galaxy in B0712+472. The final catalog contains 986 objects. The
approximate limiting magnitudes are $B \sim 25.5$, $R \sim 24.5$, and
$I \sim 23.5$ for a 5-$\sigma$ detection in our standard
aperture. Because of the smaller field-of-view of LRIS compared to the
Palomar 60-in camera, some of our spectroscopic targets (see \S3.2) do
not have available LRIS imaging.  Those for which LRIS imaging is
available are listed in Table 1; those without LRIS imaging are listed
separately in Table 2 (see \S4).

\subsection{Spectroscopic}

The spectroscopic observations were performed on March 28, 2000 and
January 17-19, 2001 with LRIS on the Keck I telescope. Multi-slit and
long-slit observations of galaxies in the B0712+472 field were made
with LRIS in spectroscopic mode using an $300~{\rm g~mm^{-1}}$ grating
blazed at 5000 \AA. The chosen grating provided a dispersion of 2.44
\AA\ per pixel and a spectral coverage of 5100 \AA. The grating angle
was set in order to provide coverage from approximately 4400 \AA\ to
9500 \AA. In order to obtain the full wavelength range along the
dispersion axis, the field-of-view of the multislit spectral
observations was reduced to approximately $2\arcmin \times
8\arcmin$. Conditions during the March 2000 observations consisted of
light to moderate cirrus and marginal seeing which ranged from
1\farcs{0} to 1\farcs{5}. The weather during the January 2001
observations was predominately clear with only thin cirrus at times.
The seeing ranged from 0\farcs{7} to 1\farcs{0}.

Spectroscopic candidates were chosen from the Palomar $gri$ imaging.
First, we included only those objects which were classified as
galaxies by SExtractor \citep[a star--galaxy classification of $< 0.8$; 
see][]{sextractor}.  Secondly, we wanted to target specifically
those galaxies which were most likely associated with the lens and,
therefore, at the same redshift.  To achieve this, we selected
galaxies which had similar colors, $(r-i) \pm 0.2$, to the lensing
galaxy. The measured color of the lensing galaxy from the Palomar
60-in imaging is $(r-i) = 0.45$. This color is consistent with the
$(F814W-F555W)$ color measured from the existing WFPC2 images (see \S
2) and with a no-evolution model of an elliptical galaxy at the lens
redshift \citep[e.g.,][]{fukugita}.

Two slitmasks were made for this lens field.  In selecting objects for
the slitmask, those galaxies that had similar colors and were
spatially close to the lensing galaxy were weighted most
heavily. Galaxies which were bluer or redder than our original color
cut were also included (though with lower weights) in order to
optimize the total number of objects per mask.  A slit width of
1\farcs{0} and a minimum slit length of 20\arcsec\ were used.  In the
end, we obtain 25 and 29 objects, respectively, on the two masks. The
first mask was observed during the March 2000 run. Two exposures of
equal duration were taken with a total exposure time of 2400 sec. The
second slitmask was observed during the January 2001 run with 
three exposures of equal duration for a total exposure time of 5400
sec.

In addition to the slitmasks, we have also used the 1\farcs{0}
long-slit, which is $175''$ in length, to observe galaxies which were
very close to the lensing galaxy and which were not included on either
of the two slitmasks.  Three long-slit positions were observed, each
covering two galaxies. Two exposures of equal duration were taken at
each position. The total exposure times were 2400, 3600, and 2400
seconds for the first, second, and third long-slit position,
respectively.

To calibrate both the multi-slit and long-slit observations,
flat-fielding and wavelength calibration were performed using internal
flat-field and arc lamp exposures which were taken after each science
exposure. Observations of the \citet{okestd} spectrophotometric
standard stars Feige 34 and G191B2B were used to remove the response
function of the chip.

\section{Results}

In Tables 1 \& 2, we present the observed parameters, including
distance from the lensing galaxy, the magnitudes and colors from
either the Keck 10-m or Palomar 60-in observations, and the measured
redshift, for all galaxies which were spectroscopically observed in
the B0712+472 field.  The redshift uncertainties have been estimated
by taking the RMS scatter in the redshifts calculated from the
individual spectral lines.

\begin{figure}
\plotone{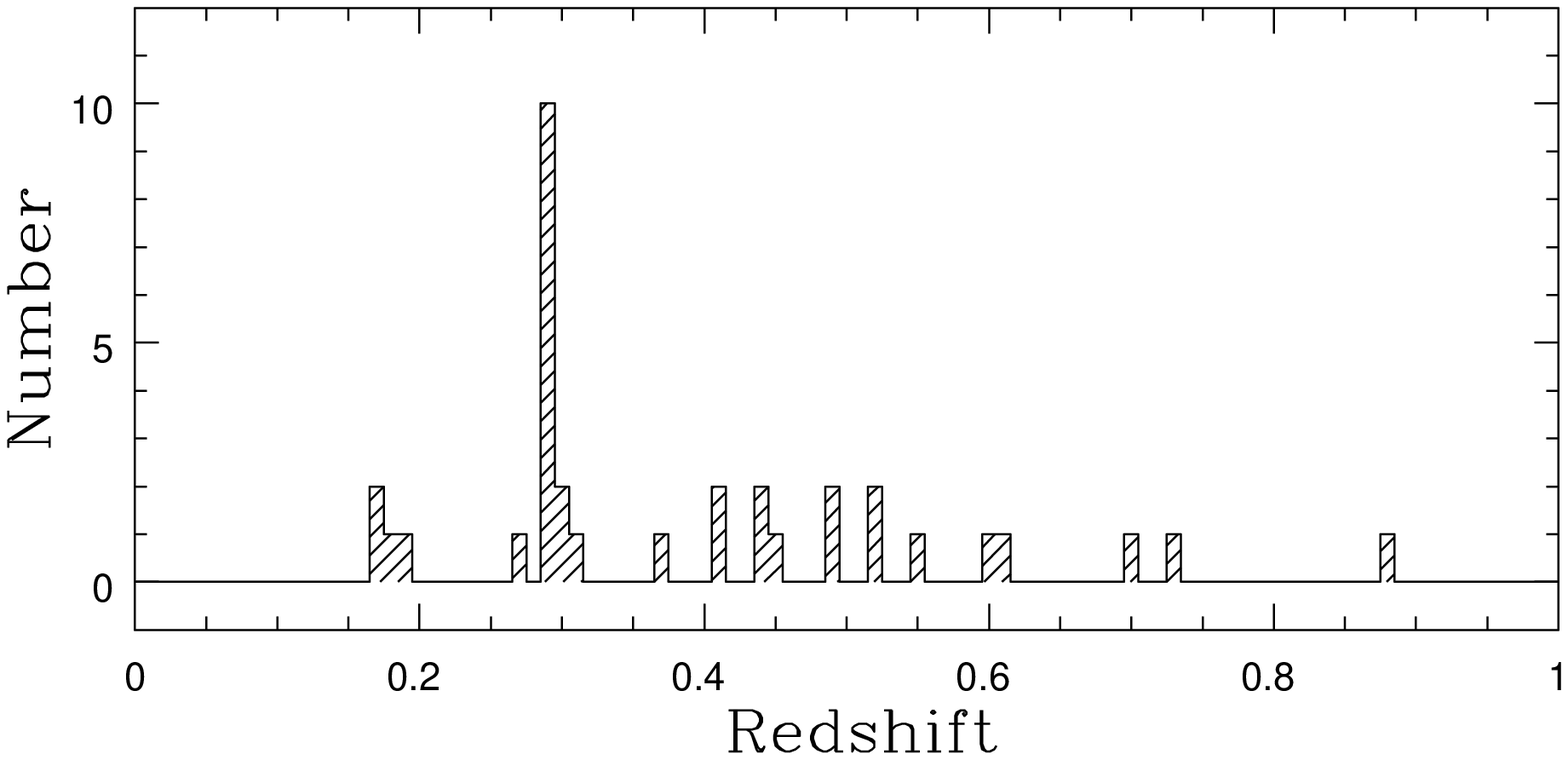}
\caption{
Histogram of redshifts (excluding Galactic stars) obtained in
the B0712+472 field.  The clear peak in the redshift distribution at
$z \sim 0.3$ indicates the presence of a real system of galaxies.
\label{fig_zhist}}
\end{figure}

In Figure~\ref{fig_zhist}, we plot the resulting distribution of redshifts
(excluding Galactic stars) in the B0712+472 field. The redshift
distribution shows a peaky structure which is typical of larger pencil
beam surveys \citep[e.g.,][]{cohen96}.  In addition, there is one
significant peak in the distribution.  While we have not detected a
group of galaxies associated with the lensing galaxy at $z_\ell =
0.4060$, we have detected a foreground group, consisting of ten
galaxies with a mean redshift of $\bar{z} = 0.2909$.

\subsection{Global Group Properties}

\begin{figure}
\plotone{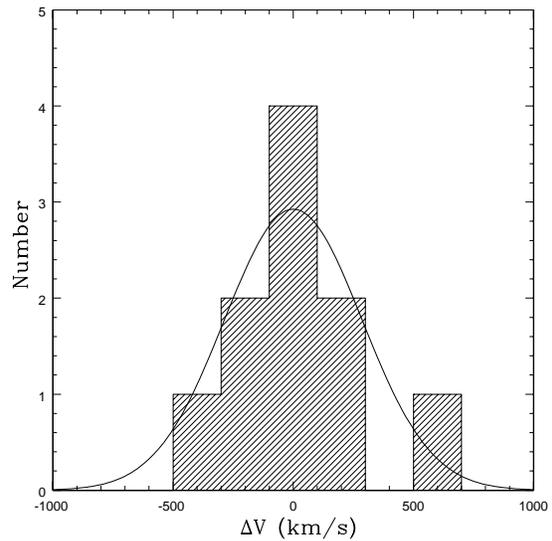}
\caption{
Histogram of the relativistically corrected velocity offsets
for foreground group at $\bar{z} = 0.2909$. Offsets are relative to
the mean group redshift. The best-fit Gaussian distribution is shown
for comparison.
\label{fig_vhist}}
\end{figure}

Using the redshifts of the ten group members, we measure a group
velocity dispersion (corrected for cosmological effects) of $\sigma =
306^{+110}_{-58}$ km s$^{-1}$.  This velocity dispersion has also been
corrected for redshift measurement errors which are typically about
100 km s$^{-1}$ (see Table 1).  We compute the uncertainty in the
dispersion according to the prescription of \citet{danese80}, which
assumes that the errors in velocity dispersions can be modeled as a
$\chi^2$ distribution and that a galaxy's velocity deviation from the
mean group redshift is independent of the galaxy's mass (i.e., the
group is virialized).  For more details, see \S3 of \citet{P98}.
Figure~\ref{fig_vhist} shows the histogram of velocity offsets for the
groups members and the best-fit Gaussian to the distribution.

\begin{figure}
\plotone{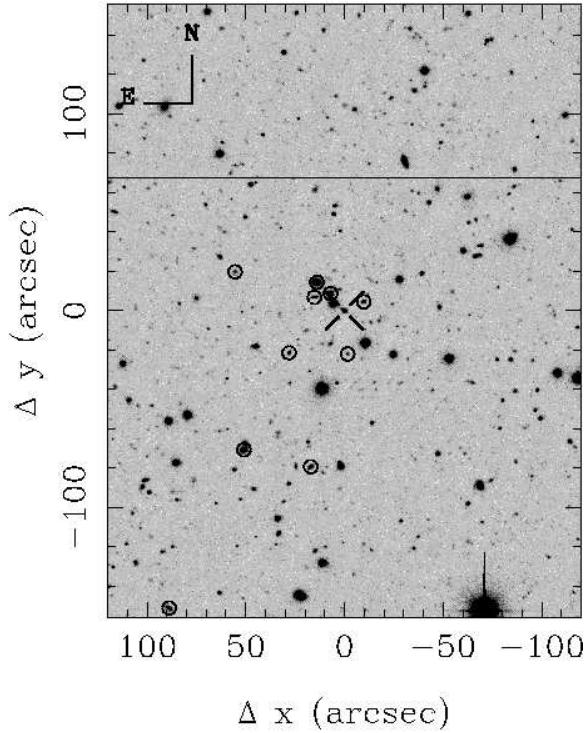}
\caption{
Composite Keck $R$-band image of the B0712+472 field. The
cross-hairs indicate the position of the lensing galaxy at $z =
0.4060$.  The ten members of the foreground group at $\bar{z} =
0.2909$ are circled.  The horizontal line indicates the northern limit
of the Palomar 60-in images. The field-of-view is $4.0\arcmin \times 
5.2\arcmin$.
\label{fig_keckim}}
\end{figure}

Figure~\ref{fig_keckim} shows the position of these galaxies, relative
to the lensing galaxy in B0712+472, on the composite $R$ band
image. The spatial distribution of the group galaxies is reasonably
compact with a harmonic radius of $R_h = 0.18 \pm 0.02~h^{-1}$ Mpc
and is centered close to the gravitational lens system. The radius and
velocity dispersion of the group imply a virial mass of
$3.0^{+2.2}_{-1.2} \times 10^{13}~h^{-1} M_{\odot}$ and a crossing
time (in units of the Hubble time) of $t_c/t_0 = 0.02$. This short
crossing time is indicative of a bound system of galaxies, such as a
group or rich cluster \citep{ramella89}.

The properties of this distant group are completely consistent with
the range of values found in nearby groups of galaxies
\citep[e.g.,][]{geller83,tully87,ramella89,zm98}.  For example,
\citet{ramella89} have studied 92 galaxy groups at $cz \le 15,000$ km
s$^{-1}$.  For the 36 rich groups which contain 5 or more members,
they find that their mean properties are $\sigma = 228 \pm 147$ km
s$^{-1}$, $R_h = 0.52 \pm 0.34~h^{-1}$ Mpc, and $t_c/t_0 = 0.06 \pm
0.05$, respectively. We have estimated the dispersion on each
parameter as $\Delta = 0.741 \times {\rm IQR}$, where IQR is the
interquartile range listed in Table 6 of \citet{ramella89}.

\subsection{Properties of the Group Galaxies}

\begin{figure}
\plotone{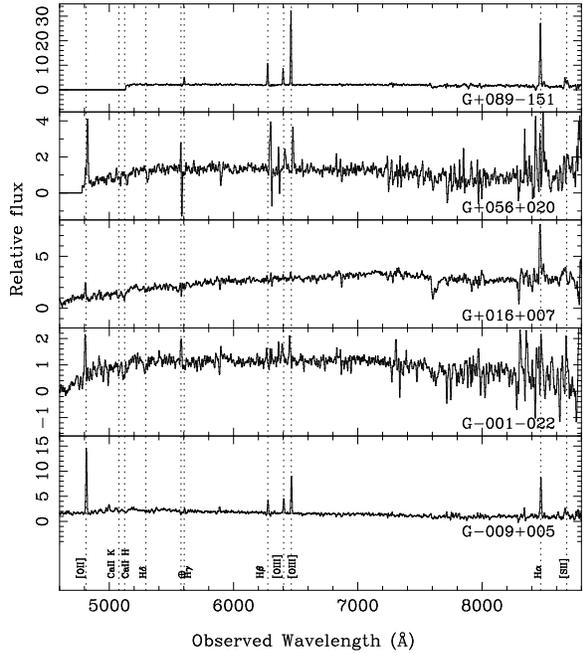}
\caption{
LRIS spectra of the five confirmed group members which are
active, star-forming galaxies. Each spectrum has been smoothed with a
box car of size 7\AA.  The vertical dashed lines represent the
location of either terrestrial atmospheric features ($\oplus$) or
typical galaxy spectral lines redshifted to the mean group redshift of
$\bar{z} = 0.2909$.
\label{fig_starform}}
\end{figure}

\begin{figure}
\plotone{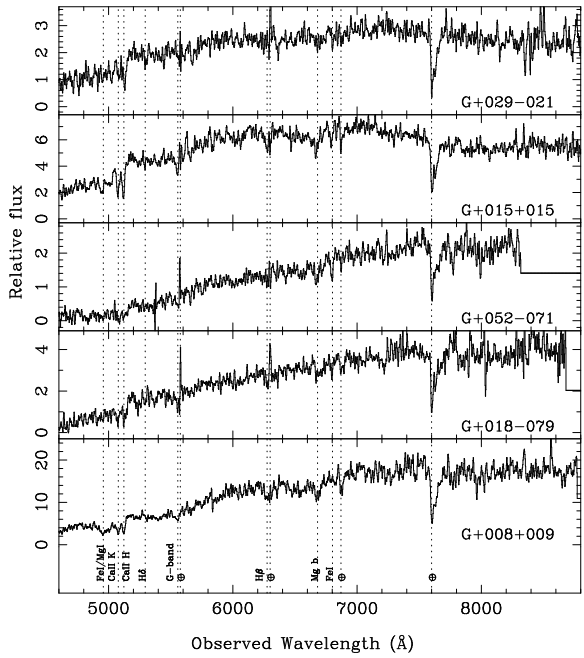}
\caption{
LRIS spectra of the five confirmed group members which are
elliptical-like galaxies.  Each spectrum has been smoothed with a box
car of size 7\AA.  The \ion{Na}{1}~D absorbtion feature is obscured by
the atmospheric A-band.
\label{fig_oldstar}}
\end{figure}

Figures \ref{fig_starform} and \ref{fig_oldstar} show the spectra of
the ten member galaxies. Five of the ten galaxies are active,
star-forming galaxies (Figure~\ref{fig_starform}). Their spectra are
characterized by strong emission lines, including [\ion{O}{2}],
[\ion{O}{3}], $H_{\alpha}$, and/or $H_{\beta}$. The rest-frame
[\ion{O}{2}] (or $H_{\alpha}$ in the case of galaxy G+016+007)
equivalent widths are all in excess of 20\AA. The remaining five
galaxies are elliptical-like in nature
(Figure~\ref{fig_oldstar}). They exhibit classic K-star absorption
features, including \ion{Ca}{2} H \& K, and/or G-band
absorption. These spectra show little or no [\ion{O}{2}] emission with
rest-frame equivalent widths of less than 15\AA.

\begin{figure}
\plotone{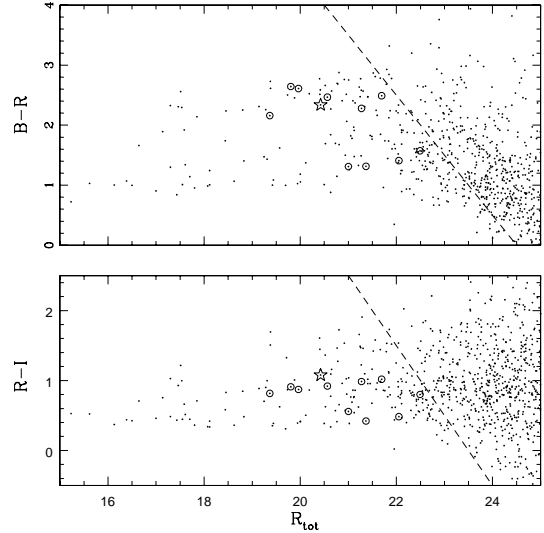}
\caption{
Color-magnitude diagrams derived from the Keck $BRI$ imaging
of the B0712+472 field.  The galaxy colors, calculated from magnitudes
measured within an aperture of radius 3\arcsec, are plotted against
the total $R$-band magnitude, $R_{tot}$, as measured by the SExtractor
package \citep{sextractor}. The lensing galaxy at $z =
0.4060$ is indicated by a star.  The galaxies which are
spectroscopically confirmed members of the foreground group of
galaxies at $\bar{z} = 0.2909$ are circled.  The magnitude limits of
the Keck imaging are indicated by the dashed lines.
\label{fig_CM}}
\end{figure}

In Figure~\ref{fig_CM}, we plot the color-magnitude diagrams (CMDs)
obtained from the Keck $BRI$ imaging of the field containing
B0712+472. The lensing galaxy and the ten galaxies which are members
of the foreground group are indicated. In the $(B-R)$ versus $R_{tot}$
CMD there is a reasonably well-defined sequence of galaxies which is
redder ($B-R \sim 2.5$) than the typical blue, field galaxy ($B-R \sim
1.0 - 1.8$). This red sequence includes the lensing galaxy at $z_\ell
= 0.4060$, as well as the five elliptical-like galaxies in the
foreground group at $\bar{z} = 0.2909$. The average colors of the five
foreground galaxies are $\langle B-R \rangle = 2.47 \pm 0.08$ and
$\langle R-I \rangle = 0.91 \pm 0.03$.

Of the remaining five group members, four are significantly bluer, as
expected from their star-forming spectral features. Their average
colors are $\langle B-R \rangle = 1.40 \pm 0.06$ and $\langle R-I
\rangle = 0.57 \pm 0.08$. The final galaxy classified as star-forming
is G+016+007. Although its spectrum exhibits \ion{Ca}{2} H \& K
absorption, it also contains relatively strong [\ion{O}{2}] and
$H_{\alpha}$ emission (see Figure~\ref{fig_starform}) with rest-frame
equivalent widths of $\sim 14$ and 23\AA, respectively. The galaxy's
colors, however, are noticeably redder than the other four active
galaxies. Here, $(B-R) = 2.28$ and $(R-I) = 0.99$. These results may
indicate a galaxy with a relatively old stellar population which has
recently undergone a burst of star formation due to an interaction or
merger.

Local galaxy groups follow a well-established correlation between
early-type fraction and velocity dispersion \citep{hickson,zm98}.  As
the velocity dispersion increases from 100 to 450 km s$^{-1}$, the
early-type fraction increases from effectively zero to 55\%. These
measures include all ($\sim 20-50$) member galaxies that are brighter
than $M_B \sim -16$ to $-17 + 5~{\rm log}~h$ and are within a radius
of $\sim 0.6-0.8 h^{-1}$ Mpc \citep{zm98}.  Based on the local
correlation, we expect the early-type fraction for our
moderate-redshift group to be approximately $30^{+20}_{-10}\%$.  From
the spectroscopic data, we find that five (out of ten) galaxies have
spectra which are typical of early-type galaxies, while the remaining
five galaxies are star-forming, implying that they are most likely
late-type (spiral or irregular) galaxies. This suggests an early-type
fraction of 50\%, which is on the high end of that expected from the
local correlation.  This result, however, may not be statistically
significant because the scatter in the local correlation is reasonably
large \citep[see Figure 7 of][]{zm98}. More importantly, our survey
does not reach the faint magnitudes of the local observations.  If we
apply a magnitude limit more typical of our survey ($M_B \sim -18 +
5~{\rm log}~h$) to the local group data, we see an increase in the
early-type fraction associated with a given velocity dispersion,
although the scatter in the relationship also increases.  This is not
unexpected since the brightest galaxies in groups are normally
early-types.  For the groups with the highest velocity dispersion, the
early-type fraction approaches 80\% or more.

To measure accurately the early-type fraction, we need visual
classifications of the group galaxies. Fortunately, we can measure
these morphologies for all but one of the group members through the
publically-available HST images.  The WFPC2 observations of B0712+472
were taken as part of two separate programs, GO-5908 (PI Jackson) and
GO-9133 (PI Falco, as part of the CASTLES program). We have chosen
to analyze the GO-9133 observations because the exposure times are
longer, the images have been dithered in three positions to improve
the angular resolution, and the position angle provides a more
beneficial field-of-view.  The observations were taken in the F555W
and F814W filters for a total exposure time of 2200 seconds each. We
have reduced these data using the ``drizzle'' software \citep{drizzle}
given in the STSDAS package {\bf ditherII}.  Nine of the ten group
members fall within the WFPC2 field-of-view. We have used data in the
reddest band F814W to make a visual classification of these galaxies
based on the Revised Hubble scheme
\citep[e.g.,][]{hubatlas,ociwatlas}. The classifications \citep[and
other information; for details, see][]{smail97,lml98} are listed in
Table 3.

\begin{figure}
\plotone{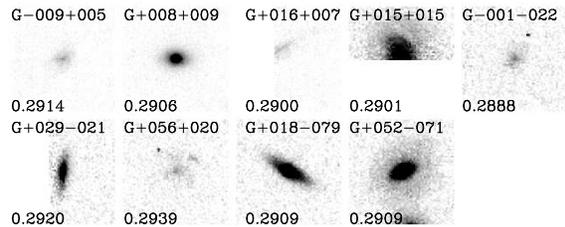}
\caption{
The nine group members in the composite F814W WFPC2 image of
the B0712+472 field.  The field-of-view of each panel is approximately
$6\arcsec \times 6\arcsec$. The galaxy identification number and the redshift
are given in the upper and lower left corner of each panel,
respectively. The galaxies are ordered by distance from the lens
system.  The orientation of each image is set by the WFPC2 rotation angle
at the time of the observation plus an additional chip-dependent term.
\label{fig_wfpc2_poststamp}}
\end{figure}

Figure~\ref{fig_wfpc2_poststamp} shows the postage stamp images of
each galaxy. The four galaxies which have very strong [\ion{O}{2}]
and/or $H_{\alpha}$ emission (G--009+005, G+016+007, G--001--022,
G+056+020; see Figure~\ref{fig_starform}) are classified as late-type
spiral galaxies. The remaining five galaxies which have
elliptical-like spectra, with little or no [\ion{O}{2}] emission
(G+008+009, G+015+015, G+029--021, G+018--079, G+052--071; see
Figure~\ref{fig_oldstar}), are classified as either elliptical, S0 or
Sa galaxies.  The only galaxy for which we do not have a visual
classification, G+089--151, is the one which is farthest from the lens
system.  This galaxy has extremely strong $H_{\alpha}$, $H_{\beta}$,
and [\ion{O}{3}] emission and is, therefore, almost certainly a
late-type (spiral or irregular) galaxy. Based on this evidence, we
conclude that the early-type fraction is 40\% and that the
morphological content of this moderate-redshift group is consistent
(within the uncertainties) with that of nearby groups of galaxies.

\section{Effect on Gravitational Potential}

The presence of a group of galaxies centered close to a
gravitational lens system will perturb the lensing gravitational
potential and affect the lensed images of the background source.  The
effect of the group mass distribution on the main gravitational
potential may be expressed in terms of an additional convergence
($\kappa_T(\vec{r})$) and shear ($\gamma_T(\vec{r})$).  The
convergence describes the isotropic component of the magnification of
an image produced at a position $\vec{r}$, while the shear describes
the anisotropic distortion of the image.  Some lens systems require a
strong shear component in order to produce the observed image
configuration.  Particularly notable is the JVAS B1422+231 system, in
which the early lens models required a strong external shear
\citep[e.g.,][]{hogg1422,keetonshear}, possibly in addition to an
extremely flattened mass distribution in the primary lensing galaxy
\citep{kormann94}.  Later observations found a group of galaxies that
provided the necessary external shear \citep{kundic1422}.  A simple
singular isothermal sphere (SIS) plus external shear model for the
CLASS B0712+472 system suggests that the external shear in this system
should be small \citep[$0.05\pm0.04$;][]{keetongalevol}.  We note that
this model is not a particularly good fit to the observations
\citep[$\chi^2/N_{\rm dof} = 4.7$;][]{keetongalevol} and might be
improved by a multiple-shear model \citep[e.g.,][]{keetonshear}.
However, it is probably correct in indicating that a very large
external shear is not needed for this system.

We can estimate the shear contribution of the foreground group to
CLASS B0712+472 and compare it to the predicted value.  We make the
approximation that the group gravitational potential is that produced
by a SIS.  The SIS has a particularly simple analytic expression for
$\gamma_T$, namely,
$$
\gamma_T(r) = \frac{b}{2r},
$$
where $b$ is the critical radius of the SIS and $r$ is the distance
from the center of the SIS, both expressed in angular units.  For a
SIS with one-dimensional velocity dispersion, $\sigma_v$, lensing a
background source at a redshift $z_s$, the critical radius (Einstein
ring radius) is
$$
b = 4\pi \frac{\sigma_v^2}{c^2} \frac{D_{\ell s}}{D_s} \approx
2.59 \left ( \frac{\sigma_v}{300~{\rm km\ s}^{-1}} \right )^2 
\frac{D_{\ell s}}{D_s}\quad {\rm arcsec}.
$$
Here, $D_{\ell s}$ and $D_s$ are the angular diameter distances from
the SIS to the background source and from the observer to the
background source, respectively.  The ratio of these distances is a
function of the redshifts of the SIS and the background source and has
a weak dependence on the cosmological model given by $\Omega_m$ and
$\Omega_\Lambda$.  For the group of galaxies discussed in this paper,
the velocity dispersion is $\sigma_v = 306~{\rm km\ s}^{-1}$ and the
redshifts are $z_{\rm SIS} = 0.2909$ and $z_s = 1.34$ for the group
and the background source, respectively.  To explore the dependence on
the cosmological model, we calculate $b$ for three common models:
$(\Omega_m, \Omega_\Lambda)$ = (1.0,0.0), (0.2,0.0), and (0.2,0.8).
The resulting critical radii are $b =$1\farcs76, 1\farcs77,
and 1\farcs97, respectively.  The convergence and shear produced by
the group at the position of the lens system, for the three
cosmologies considered, are:
$$
\kappa_T(r_\ell) = \gamma_T(r_\ell) \sim \frac{0.95}{r_\ell},
$$
where $r_\ell$ is the angular distance from the group center to the
lens system, in arcseconds.  

The effect of the foreground group on the gravitational lensing
potential depends critically on $r_\ell$.  Unfortunately, this
quantity is not well determined because the position of the group
center is not clear.  A simple calculation of the luminosity-weighted
centroid of the group gives $r_\ell = 29\arcsec$, and thus
$\gamma_T(r_\ell) \sim 0.03$, which is consistent with the value
predicted by \citet{keetongalevol}.  However, with only ten galaxies
known to be in the group, this number is subject to the effects of
small-number statistics.  For example, by excluding the galaxy most
distant from the lens system, the luminosity-weighted group centroid
shifts significantly closer to the lens, giving $r_\ell = 18\arcsec$
and $\gamma_T(r_\ell) \sim 0.05$.  

In addition, the luminosity-weighted centroid is almost certainly
biased toward the south of the lens system.  This bias arises because
a fifth-magnitude star lies $\sim$6\arcmin\ to the north-northwest of
the lens.  Scattered light from this star strongly contaminated our
Palomar 60-in images.  In order to reduce the effect of the scattered
light to a reasonable level, we centered our images $\sim$4\arcmin\ to
the south of the lens system.  The horizontal line in
Figure~\ref{fig_keckim} indicates the northern limit of the Palomar
60-in images.  Because we had no information about galaxies north of
this line, they were not included on the slitmasks.  It is likely that
the Keck images, which are much less affected by the scattered light,
will provide good candidate group members to the north of the lens
system.  The inclusion of these galaxies could shift the group
centroid farther to the north and closer to the lens system.

A final estimate of $r_\ell$, and thus $\gamma_T$, can be obtained by
exploiting a relationship seen in nearby groups of galaxies.  In local
galaxy groups for which diffuse X-ray emission has been detected, the
brightest group galaxy (BGG) is located at the center of the group
potential \citep{zm98}.  Thus, if the CLASS B0712+472 foreground group
is virialized, the distance from the BGG to the lens system should be a
good estimator of $r_\ell$.  Our survey of the group members is far
from complete due to the northern cutoff in our galaxy selection and
the sparseness of our spectroscopic coverage.  However, for the
purposes of our rough estimate, we can assume that the brightest
galaxy known to be in the group (G+015+015, with $R = 19.6$) is the BGG.
This would imply that $r_\ell = 21$\arcsec\ and, thus, $\gamma_T(r_\ell)
\sim 0.05$.

Unless further observations significantly increase the number of
galaxies known to be in this group, it is unlikely that the optical
observations alone will adequately determine the location of the
center of the group gravitational well.  A sensitive X-ray observation
with XMM or Chandra, which would provide a measure of the total and gas
mass distribution, could give a much more accurate location for the
group centroid.

\section{Conclusion}

We present the first results from our observational program which
seeks to exploit the gravitational lens--galaxy group connection to
find and study moderate to high-redshift groups of galaxies. We have
surveyed the field surrounding the gravitational lens system B0712+472
and have found a foreground group of galaxies at $\bar{z} = 0.2909$
which is coincident with the lensing galaxy. The small group, which
contains 10 member galaxies, has dynamical and morphological
properties which are consistent with those of local groups of
galaxies. We estimate that the effect of the foreground group on the
gravitational lensing potential of B0712+472 is small. Based on the simplest
model, that of a singular isothermal sphere, the group produces an
external shear which is only a few percent.  However, the magnitude of
the shear is reasonably uncertain because it is inversely proportional
to the distance between the group center and the lens system. Based on
the relatively small number of confirmed group members, this distance
is uncertain by at least a factor of 2.

Currently, there are five spectroscopically-confirmed groups (as
opposed to massive clusters) detected by their association with
gravitational lens systems.  In Table 4, we list the properties of all
five groups, including the one presented in this paper.  Remarkably,
the redshift of each group is $z \sim 0.3$. Each is quite compact with
harmonic radii less than $0.18~h^{-1}$ Mpc.  Based on the small number
of confirmed group members, we find that the velocity dispersions
range by more than an order of magnitude, from $\sim 30 - 535$ km
s$^{-1}$.  However, the smallest velocity dispersion is highly
uncertain because the group, that associated with MG~0751+2716, has
only three confirmed members.  Further studies of these systems are
clearly required in order to make accurate measurements of the
dynamical properties of the groups, as well as the spectral and
morphological properties of their member galaxies.  Because all five
groups have very similar redshifts, they comprise an unique sample in
which to study the diversity of group properties at this redshift.

\acknowledgements

We would like to thank Marc Postman, David Rusin, Alice Shapley,
Gordon Squires, and Ann Zabludoff for helpful discussions and
essential aids to this paper.  We are grateful to Karl Dunscombe, Skip
Staples, Jean Mueller, Steve Kunsman, Dave Tennent, Greg Van Idsinga,
Ron Quick, Meg Whittle, Barbara Schaefer, Randy Campbell, Bob
Goodrich, Grant Hill, David Sprayberry, Greg Wirth, and the rest of
the staff at the Palomar and Keck Observatories for keeping the
telescopes operating smoothly.  The W. M. Keck Observatory is operated
as a scientific partnership between the California Institute of
Technology, the University of California, and the National Aeronautics
and Space Administration.  It was made possible by generous financial
support of the W. M. Keck Foundation.  Some of the data presented in
this paper were obtained from the Multimission Archive at the Space
Telescope Science Institute (MAST). STScI is operated by the
Association of Universities for Research in Astronomy, Inc., under
NASA contract NAS5-26555. Support for MAST for non-HST data is
provided by the NASA Office of Space Science via grant NAG5-7584 and
by other grants and contracts.  The National Radio Astronomy Observatory
is operated by Associated Universities, Inc., under cooperative
agreement with the National Science Foundation.  MERLIN is operated as
a National Facility by NRAL, University of Manchester, on behalf of
the UK Particle Physics and Astronomy Research Council.


\newpage

\newpage



\begin{center}
\begin{deluxetable}{crrrccccrrc}
\rotate
\tablewidth{0pt}
\tabletypesize{\scriptsize}
\tablenum{1}
\tablecaption{Photometric and Spectroscopic Results for the B0712+472 Field: 
Objects with Keck Imaging}
\tablehead{
\colhead{Galaxy ID\tablenotemark{a}} &
\colhead{$X$} &
\colhead{$Y$} &
\colhead{$\Delta r$(\arcsec)\tablenotemark{b}} &
\colhead{$B_{tot} \pm \Delta B_{tot}$\tablenotemark{c}}    &
\colhead{$R_{tot} \pm \Delta R_{tot}$\tablenotemark{c}}    &
\colhead{$I_{tot} \pm \Delta I_{tot}$\tablenotemark{c}}    &
\colhead{$R_{ap}$}    &
\multicolumn{1}{c}{${(B-R)}_{ap}$} &
\multicolumn{1}{c}{${(R-I)}_{ap}$} &
\colhead{$z \pm \Delta z$}}
\startdata
G$+$103$-$090 &  90.19 & 504.37 &  136.8 & $22.61 \pm 0.01$\tablenotemark{d} &  $21.12 \pm 0.01$\tablenotemark{d} &  $20.59 \pm 0.01$\tablenotemark{d} &  21.13\tablenotemark{d}&   1.477 &  0.540 & $0.3041 \pm 0.0004$ \\
G$+$093$-$073 & 140.86 & 582.49 &  118.1 & \nodata          &  $23.48 \pm 0.05$ &  $21.68 \pm 0.01$ &  23.45& \nodata &  1.812 & $0.8838 \pm 0.0004$ \\
G$+$089$-$151 & 155.10 & 212.98 &  175.6 & $22.32 \pm 0.01$ &  $21.00 \pm 0.01$ &  $20.44 \pm 0.01$ &  21.04&   1.311 &  0.557 & $0.2905 \pm 0.0002$ \\
G$+$075$-$183 & 221.68 &  60.80 &  198.0 & $23.83 \pm 0.02$ &  $22.63 \pm 0.02$ &  $21.86 \pm 0.01$ &  22.58&   1.264 &  0.721 & $0.4917 \pm 0.0002$ \\
G$+$066$-$098 & 267.65 & 465.08 &  118.1 & $25.33 \pm 0.08$ &  $22.61 \pm 0.01$ &  $20.77 \pm 0.01$ &  22.58&   2.651 &  1.854 & $0.0000$ \\
G$+$056$+$020 & 313.60 &1022.05 &   59.7 & $23.46 \pm 0.02$ &  $22.05 \pm 0.01$ &  $21.57 \pm 0.01$ &  22.05&   1.410 &  0.483 & $0.2939 \pm 0.0002$ \\
G$+$052$-$071 & 334.83 & 593.45 &   87.5 & $22.61 \pm 0.01$\tablenotemark{d} &  $19.96 \pm 0.01$\tablenotemark{d} &  $19.08 \pm 0.01$\tablenotemark{d} &  19.97\tablenotemark{d}&   2.611 &  0.872 & $0.2909 \pm 0.0004$ \\
G$+$051$-$068 & 338.10 & 608.25 &   84.8 & $22.44 \pm 0.01$\tablenotemark{d} &  $20.81 \pm 0.01$\tablenotemark{d} &  $20.18 \pm 0.01$\tablenotemark{d} &  20.78\tablenotemark{d}&   1.677 &  0.654 & $0.2687 \pm 0.0004$ \\
G$+$047$-$091 & 358.42 & 498.97 &  102.0 & $23.83 \pm 0.02$ &  $21.40 \pm 0.01$ &  $20.44 \pm 0.01$ &  21.38&   2.476 &  0.956 & $0.4053 \pm 0.0003$ \\
G$+$044$-$167 & 367.50 & 135.59 &  173.3 & $22.54 \pm 0.01$ &  $21.21 \pm 0.01$\tablenotemark{d} &  $20.59 \pm 0.01$\tablenotemark{d} &  21.21\tablenotemark{d}&   1.331 &  0.621 & $0.1742 \pm 0.0003$ \\
G$+$041$-$181 & 384.33 &  70.38 &  185.6 & $20.62 \pm 0.01$\tablenotemark{d} &  $19.25 \pm 0.01$\tablenotemark{d} &  $18.56 \pm 0.01$\tablenotemark{d} &  19.59\tablenotemark{d}&   1.473 &  0.703 & $0.1758 \pm 0.0002$ \\
G$+$041$-$159 & 385.08 & 174.10 &  164.3 & $24.11 \pm 0.04$ &  $22.14 \pm 0.01$ &  $20.97 \pm 0.01$ &  22.15&   1.914 &  1.183 & $0.6165 \pm 0.0002$ \\
G$+$034$-$064 & 415.75 & 624.91 &   72.4 & $24.86 \pm 0.05$\tablenotemark{d} &  $23.15 \pm 0.03$\tablenotemark{d} &  $22.59 \pm 0.03$\tablenotemark{d} &  23.00\tablenotemark{d}&   1.512 &  0.576 & $0.3673 \pm 0.0003$ \\
G$+$034$-$113 & 417.71 & 392.75 &  118.0 & $23.09 \pm 0.01$ &  $21.11 \pm 0.01$ &  $20.35 \pm 0.01$\tablenotemark{d} &  21.10&   1.967 &  0.759 & $0.3041 \pm 0.0012$ \\
G$+$029$-$021 & 443.14 & 827.18 &   35.8 & $24.21 \pm 0.03$ &  $21.69 \pm 0.01$ &  $20.67 \pm 0.01$ &  21.68&   2.490 &  1.018 & $0.2920 \pm 0.0001$ \\
G$+$020$-$010 & 483.40 & 880.76 &   22.6 & $24.79 \pm 0.04$ &  $22.72 \pm 0.02$ &  $21.26 \pm 0.01$ &  22.68&   2.089 &  1.433 & \nodata \\
G$+$018$-$031 & 492.29 & 782.94 &   35.7 & $25.30 \pm 0.09$\tablenotemark{d} &  $23.09 \pm 0.03$\tablenotemark{d} &  $21.07 \pm 0.01$\tablenotemark{d} &  23.09\tablenotemark{d}&   2.180 &  1.657 & $0.8116 \pm 0.0001$ \\
G$+$018$-$079 & 494.99 & 553.14 &   81.2 & $23.00 \pm 0.01$ &  $20.57 \pm 0.01$ &  $19.64 \pm 0.01$ &  20.56&   2.467 &  0.920 & $0.2909 \pm 0.0005$ \\
G$+$015$-$122 & 505.21 & 350.79 &  123.0 & $23.83 \pm 0.03$ &  $21.41 \pm 0.01$ &  $20.29 \pm 0.01$\tablenotemark{d} &  21.39&   2.471 &  1.018 & $0.4505 \pm 0.0004$ \\
G$+$016$+$007 & 504.10 & 960.50 &   17.4 & $23.58 \pm 0.02$ &  $21.27 \pm 0.01$\tablenotemark{d} &  $20.23 \pm 0.01$\tablenotemark{d} &  21.37\tablenotemark{d}&   2.280 &  0.986 & $0.2900 \pm 0.0001$ \\
G$+$015$+$015 & 509.65 & 997.07 &   20.9 & $21.56 \pm 0.01$\tablenotemark{d} &  $19.37 \pm 0.01$\tablenotemark{d} &  $18.52 \pm 0.01$\tablenotemark{d} &  19.50\tablenotemark{d}&   2.160 &  0.816 & $0.2901 \pm 0.0002$ \\
G$+$010$-$120 & 530.96 & 360.63 &  120.4 & $25.54 \pm 0.07$ &  $22.89 \pm 0.03$ &  $21.87 \pm 0.02$ &  22.89&   3.758 &  1.035 & $0.4530 \pm 0.0003$ \\
G$+$008$+$009 & 542.70 & 969.82 &   11.9 & $22.45 \pm 0.01$\tablenotemark{d} &  $19.80 \pm 0.01$\tablenotemark{d} &  $18.89 \pm 0.01$\tablenotemark{d} &  19.83\tablenotemark{d}&   2.644 &  0.910 & $0.2906 \pm 0.0002$ \\
G$+$007$-$017 & 543.30 & 849.15 &   18.2 & $25.14 \pm 0.07$\tablenotemark{d} &  $21.85 \pm 0.01$\tablenotemark{d} &  $20.09 \pm 0.01$\tablenotemark{d} &  21.86\tablenotemark{d}&   3.319 &  1.730 & $0.8833 \pm 0.0001$ \\
G$+$006$+$050 & 550.69 &1161.25 &   50.0 & $23.49 \pm 0.01$ &  $20.78 \pm 0.01$\tablenotemark{d} &  $19.51 \pm 0.01$\tablenotemark{d} &  20.77\tablenotemark{d}&   2.687 &  1.274 & $0.0000$ \\
G$+$005$+$054 & 554.18 &1181.12 &   54.0 & $23.77 \pm 0.02$ &  $21.60 \pm 0.01$\tablenotemark{d} &  $20.64 \pm 0.01$\tablenotemark{d} &  21.59\tablenotemark{d}&   2.164 &  0.965 & \nodata \\
G$+$005$-$037 & 558.10 & 754.35 &   36.9 & $22.61 \pm 0.01$\tablenotemark{d} &  $21.73 \pm 0.01$\tablenotemark{d} &  $20.70 \pm 0.01$\tablenotemark{d} &  21.72\tablenotemark{d}&   0.961 &  0.859 & \nodata \\
G$+$003$+$041 & 566.08 &1120.59 &   41.1 & $24.47 \pm 0.03$ &  $22.32 \pm 0.01$ &  $21.53 \pm 0.01$ &  22.32&   2.212 &  0.788 & $0.0000$ \\
G$-$001$-$022 & 584.12 & 824.32 &   21.8 & $24.08 \pm 0.03$ &  $22.49 \pm 0.02$ &  $21.75 \pm 0.01$\tablenotemark{d} &  22.47&   1.573 &  0.800 & $0.2888 \pm 0.0001$ \\
G$-$009$+$005 & 622.75 & 949.82 &   10.5 & $22.67 \pm 0.01$ &  $21.37 \pm 0.01$ &  $20.92 \pm 0.01$ &  21.34&   1.315 &  0.421 & $0.2914 \pm 0.0001$ \\
G$-$013$-$006 & 639.48 & 900.34 &   14.0 & $24.62 \pm 0.04$ &  $22.91 \pm 0.02$ &  $21.44 \pm 0.01$\tablenotemark{d} &  23.01&   1.769 &  1.562 & \nodata \\
G$-$028$-$107 & 711.75 & 422.28 &  110.5 & $24.44 \pm 0.03$ &  $21.66 \pm 0.01$ &  $20.24 \pm 0.01$ &  21.65&   2.783 &  1.441 & $0.0000$ \\
G$-$047$+$062 & 798.78 &1221.65 &   77.8 & $23.80 \pm 0.03$ &  $21.01 \pm 0.01$ &  $19.94 \pm 0.01$\tablenotemark{d} &  21.06&   2.752 &  1.038 & $0.4369 \pm 0.0001$ \\
G$-$063$+$045 & 873.53 &1140.66 &   77.2 & $24.10 \pm 0.03$ &  $22.08 \pm 0.01$\tablenotemark{d} &  $21.32 \pm 0.01$ &  22.04\tablenotemark{d}&   2.003 &  0.750 & \nodata \\

\tablenotetext{a}{The two numbers which comprise the galaxy ID
represent the distance in RA($''$) and Dec($''$) from the lensing
galaxy.}

\tablenotetext{b}{The distance from the lensing galaxy in arcseconds.}

\tablenotetext{c}{Variable-diameter aperture magnitude measured in an
 elliptical aperture of major axis radius of $2.5 \times r_k$ (where
 $r_{k}$ is the Kron radius), unless an error flag occurred in which
 case the ``corrected'' isophotal magnitude is used (see Bertin \&
 Arnouts 1996).}

\tablenotetext{d}{An error flag occurred indicating that bright
neighbors may bias magnitude estimate and/or it was originally a
blend.}

\enddata
\end{deluxetable}
\end{center}


\begin{deluxetable}{crrrccccrrc}
\rotate
\tablewidth{0pt}
\tabletypesize{\scriptsize}
\tablenum{2}
\tablecaption{Photometric and Spectroscopic Results for the B0712+472 Field: 
Objects with Palomar 60-in Imaging}
\tablehead{
\colhead{Galaxy ID\tablenotemark{a}} &
\colhead{$X$} &
\colhead{$Y$} &
\colhead{$\Delta r$(\arcsec)\tablenotemark{b}} &
\colhead{$g_{tot} \pm \Delta g_{tot}$\tablenotemark{c}}    &
\colhead{$r_{tot} \pm \Delta r_{tot}$\tablenotemark{c}}    &
\colhead{$i_{tot} \pm \Delta i_{tot}$\tablenotemark{c}}    &
\colhead{$r_{ap}$}    &
\multicolumn{1}{c}{${(g-r)}_{ap}$} &
\multicolumn{1}{c}{${(r-i)}_{ap}$} &
\colhead{$z \pm \Delta z$}}
\startdata
G$+$102$-$338&  661.77&  489.52& 352.8 & \nodata         & $22.89\pm 0.08$\tablenotemark{d} &  $22.24 \pm   0.08$\tablenotemark{d} &  22.74\tablenotemark{d} & \nodata&   0.344& \nodata \\
G$+$060$-$318&  772.06&  540.72& 323.8 & \nodata         & $22.38\pm 0.05$ &  $21.95 \pm   0.07$ &  22.35& \nodata&   0.517& $ 0.5453 \pm 0.0001$ \\
G$+$059$-$221&  775.20&  797.89& 228.5 & \nodata         & $22.91\pm 0.10$ &  $22.54 \pm   0.10$ &  22.87& \nodata&   0.514& $ 0.6044 \pm 0.0002$ \\
G$+$055$-$329&  784.91&  511.14& 334.0 & \nodata         & $22.81\pm 0.08$ &  $21.62 \pm   0.05$ &  22.80& \nodata&   1.140& $ 0.0000$ \\
G$+$053$-$305&  790.79&  575.14& 309.7 & $21.00\pm 0.02$ & $20.30\pm 0.01$ &  $20.03 \pm   0.02$ &  20.45&   0.788&   0.327& $ 0.3114 \pm 0.0001$ \\
G$+$041$-$353&  821.36&  450.10& 354.9 & $22.22\pm 0.06$ & $20.96\pm 0.02$ &  $20.49 \pm   0.03$ &  21.04&   1.430&   0.412& $ 0.5175 \pm 0.0004$ \\
G$+$041$-$203&  824.38&  845.61& 206.6 & $22.02\pm 0.04$ & $22.05\pm 0.04$ &  $21.99 \pm   0.07$ &  22.00&   0.040&   0.070& $ 0.1740 \pm 0.0004$ \\
G$+$036$-$295&  836.48&  600.64& 297.5 & $18.46\pm 0.01$\tablenotemark{d} & $17.71\pm 0.01$\tablenotemark{d} &  $17.70 \pm   0.01$\tablenotemark{d} &  17.84\tablenotemark{d}&   0.749&   0.062& $ 0.0000$ \\
G$+$024$-$232&  867.22&  768.98& 232.8 & $23.09\pm 0.07$ & $21.95\pm 0.06$ &  $22.03 \pm   0.09$ &  22.05&   1.132&  -0.003& $ 0.5173 \pm 0.0003$ \\
G$+$009$-$250&  906.15&  719.04& 250.6 & $22.36\pm 0.05$ & $21.18\pm 0.02$ &  $20.67 \pm   0.02$ &  21.18&   1.176&   0.526& $ 0.0000$ \\
G$+$008$-$327&  909.94&  516.29& 327.4 & $22.72\pm 0.07$ & $21.38\pm 0.03$ &  $21.35 \pm   0.04$ &  21.41&   1.308&   0.010& \nodata \\
G$-$003$-$348&  937.26&  462.25& 347.8 & $23.00\pm 0.09$ & $21.99\pm 0.06$ &  $21.81 \pm   0.09$ &  22.25&   0.719&   0.150& \nodata \\
G$-$009$-$206&  955.31&  835.30& 206.5 & $22.65\pm 0.07$ & $21.32\pm 0.03$ &  $20.70 \pm   0.02$ &  21.32&   1.336&   0.629& $ 0.0000$ \\
G$-$015$-$280&  969.97&  640.32& 280.6 & $23.62\pm 0.15$ & $22.87\pm 0.07$ &  $22.71 \pm   0.15$ &  23.08&   0.552&   0.535& $ 0.4931 \pm 0.0002$ \\
G$-$017$-$339&  974.33&  484.22& 339.8 & $21.64\pm 0.04$ & $20.89\pm 0.02$ &  $20.72 \pm   0.03$ &  21.00&   0.772&   0.168& $ 0.4362 \pm 0.0008$ \\
G$-$039$-$307& 1032.55&  569.32& 309.4 & $22.37\pm 0.06$ & $20.86\pm 0.02$ &  $19.74 \pm   0.01$ &  20.86&   1.461&   1.122& $ 0.0000$ \\
G$-$043$-$198& 1044.55&  856.72& 202.6 & $20.28\pm 0.01$ & $19.70\pm 0.01$ &  $19.41 \pm   0.01$ &  19.79&   0.643&   0.306& $ 0.1949 \pm 0.0002$ \\
G$-$077$-$243& 1134.94&  736.66& 255.4 & $24.00\pm 0.10$ & $22.40\pm 0.05$ &  $21.44 \pm   0.04$ &  22.37&   1.487&   0.930& $ 0.7047 \pm 0.0008$ \\
G$-$078$-$224& 1138.10&  786.81& 237.7 & $22.71\pm 0.06$ & $22.81\pm 0.06$ &  $22.95 \pm   0.11$ &  22.82&  -0.089&  -0.225& \nodata \\
G$-$087$-$262& 1159.39&  686.62& 276.3 & \nodata         & $21.72\pm 0.03$\tablenotemark{d} &  $21.32 \pm   0.04$\tablenotemark{d} &  21.53\tablenotemark{d}& \nodata&   0.361& $ 0.7269 \pm 0.0006$ \\
G$-$088$-$260& 1161.92&  692.79& 274.4 & $21.53\pm 0.03$ & $21.71\pm 0.03$\tablenotemark{d} &  $21.36 \pm   0.04$\tablenotemark{d} &  21.74\tablenotemark{d}&  -0.076&   0.344& $ 0.7260 \pm 0.0030$ \\
G$-$087$-$212& 1160.21&  819.47& 229.1 & $22.89\pm 0.06$ & $21.84\pm 0.04$ &  $21.42 \pm   0.05$ &  21.84&   0.980&   0.439& \nodata \\

\tablenotetext{a}{The two numbers which comprise the galaxy ID
represent the distance in RA($''$) and Dec($''$) from the lensing
galaxy.}

\tablenotetext{b}{The distance from the lensing galaxy in arcseconds.}

\tablenotetext{c}{Variable-diameter aperture magnitude measured in an
 elliptical aperture of major axis radius of $2.5 \times r_k$ (where
 $r_{k}$ is the Kron radius), unless an error flag occurred in which
 case the ``corrected'' isophotal magnitude is used (see Bertin \&
 Arnouts 1996).}

\tablenotetext{d}{An error flag occurred indicating that bright
neighbors may bias magnitude estimate and/or it was originally a
blend.}

\enddata
\end{deluxetable}


\begin{deluxetable}{lcrrlccl}
\scriptsize
\tablewidth{0pt}
\tabletypesize{\scriptsize}
\tablenum{3}
\tablecaption{Morphological Classification of Group Members}
\tablehead{
\colhead{Galaxy} &
\colhead{Chip} &
\colhead{$x$\tablenotemark{a}} &
\colhead{$y$\tablenotemark{a}} &
\colhead{Class\tablenotemark{b}} &
\colhead{$D$\tablenotemark{c}} &
\colhead{Interp\tablenotemark{d}} &
\colhead{Comments\tablenotemark{e}}}
\startdata
G$-$009$+$005  & 1 & 1379.24 & 1409.81 & Sc & 2 &\nodata & face-on; small bulge; diffuse, low SB disk \\
G$+$008$+$009  & 1 &  611.17 & 1315.62 & E  & 0 &\nodata & face-on \\
G$+$016$+$007  & 1 &  314.26 & 1116.14 & S  & 0 &\nodata & edge-on spiral; on edge of chip  \\
G$+$015$+$015  & 2 &  832.39 &  318.38 & Sa & 0 &\nodata & face-on; indication of arm or dust lane; on edge of chip\\
G$-$001$-$022  & 4 &  384.06 &  842.06 & Sc & 2 &\nodata & face-on; small bulge; diffuse, low SB disk \\
G$+$029$-$021  & 3 &  333.49 &  589.62 & E  & 0 &\nodata & inclined \\
G$+$056$+$020  & 2 &  625.58 & 1133.79 & Sc & 3 & C      & several bright knots in diffuse disk\\
G$+$018$-$079  & 4 & 1597.59 &  890.72 & S0 & 0 &\nodata & slightly asymmetric disk\\
G$+$052$-$071  & 3 &  406.59 & 1683.43 & S0 & 0 &\nodata & face-on; close companion to the north  \\ 
\enddata

\tablenotetext{a}{In order to combine the drizzled data, each chip is
mapped onto $2048 \times 2048$ image with a pixel scale which is
one-half of the input pixel scale. The $x$ and $y$ positions are
derived from these mapped images.}  

\tablenotetext{b}{The standard Hubble classification scheme (e.g., E, S0, 
Sa, Sab etc.).}

\tablenotetext{c}{Disturbance index : 0, normal; 1, moderate
asymmetry; 2, strong asymmetry; 3, moderate distortion; 4, strong
distortion.}

\tablenotetext{d}{Interpretation of disturbance index : M, merger; I,
tidal interaction with neighbor; T, tidal feature; C, chaotic.}

\tablenotetext{e}{Description of galaxy morphology.}

\end{deluxetable}


\begin{deluxetable}{lccrcrrrr}
\scriptsize
\tablewidth{0pt}
\tabletypesize{\scriptsize}
\tablenum{4}
\tablecaption{Spectroscopically-Confirmed Galaxy Groups Associated with
Gravitational Lenses}
\tablehead{
\colhead{Lens System} &
\colhead{$z_s$} &
\colhead{$z_{\ell}$} &
\colhead{$N_z$\tablenotemark{a}} &
\colhead{$\bar{z}$} &
\colhead{$\sigma$\tablenotemark{b}} &
\colhead{$R_h$\tablenotemark{b}} &
\colhead{$M_v$\tablenotemark{b}} &
\colhead{$t_c/t_0$\tablenotemark{b}} \\
\colhead{} &
\colhead{} &
\colhead{} &
\colhead{} &
\colhead{} &
\colhead{(km s$^{-1}$)} &
\colhead{($h^{-1}$ Mpc)} &
\colhead{($10^{13}~h^{-1}~M_{\odot}$)} &
\colhead{}}
\startdata
CLASS~B0712+472 & 1.339   & 0.4060 & 10 & 0.2909 & $306^{+110}_{-58}$   & $0.18 \pm 0.02$ & $3.0^{+2.2}_{-1.2}$ & 0.018 \\
MG~0751+2716    & 3.200   & 0.3502 &  3 & 0.3503 & \phn$27^{+56}_{-27}$ & $0.06 \pm 0.04$ & $<0.04$             & 0.060 \\
PG~1115+080     & 1.718   & 0.3098 &  4 & 0.3103 & $243^{+246}_{-84}$   & $0.06 \pm 0.01$ & $0.6^{+1.2}_{-0.4}$ & 0.007 \\
MG~1131+0456    & \nodata & 0.8440 &  3 & 0.3432 & $222^{+384}_{-65}$   & $0.14 \pm 0.01$ & $1.3^{+4.7}_{-1.1}$ & 0.017 \\
JVAS~B1422+231  & 3.626   & 0.3374 &  6 & 0.3381 & $535^{+314}_{-121}$  & $0.05 \pm 0.01$ & $2.7^{+3.2}_{-1.3}$ & 0.003 \\
\enddata

\tablenotetext{a}{Number of group members including the lensing galaxy
if applicable.}

\tablenotetext{b}{We have used the redshifts and positions given in
the papers listed in \S1 to calculate the group parameters and
errors according to the method described in \S4 and the definitions
given by \citet{ramella89}.}

\end{deluxetable}

\end{document}